\documentclass[amsmath, amssymb, preprintnumbers, showpacs,
 showkeys,aps,pra,superscriptaddress,twocolumn]{revtex4-1}

\usepackage{color} 
\usepackage{hyperref}
\usepackage{amsmath,amssymb,mathrsfs}
\usepackage{psfrag}
\usepackage{graphicx}
\usepackage{graphics}
\usepackage{subfigure}
\usepackage{epsfig}
\usepackage{bm}
\usepackage{verbatim,color,ulem}
\usepackage{braket}
\usepackage{verbatim}
\usepackage[utf8]{inputenc}
\usepackage[english]{babel}
\usepackage{amssymb}
\usepackage{amsthm}

\begin{document}
	
\title{Josephson effect with superfluid fermions in the two-dimensional 
BCS-BEC crossover}
	 
 \author{F. Pascucci}
\affiliation{Dipartimento di Fisica e Astronomia "Galileo Galilei", 
Universit\`a di Padova, via Marzolo 8, 35131 Padova, Italy}
\author{L. Salasnich}
\affiliation{Dipartimento di Fisica e Astronomia "Galileo Galilei", 
Universit\`a di Padova, via Marzolo 8, 35131 Padova, Italy}
\affiliation{Istituto Nazionale di Ottica del Consiglio Nazionale delle Ricerche, CNR-INO, 
via Nello Carrara 2, 50019 Sesto Fiorentino, Italy}
\date{\today{}}

\begin{abstract}
We investigate the macroscopic quantum tunneling of fermionic superfluids in the two-dimensional BCS-BEC 
crossover by using an effective tunneling energy which explicitly depends on the condensate fraction and the chemical 
potential of the system. We compare the mean-field effective tunneling energy with the beyond-mean-field one 
finding that the mean-field tunneling energy is not reliable in the BEC regime of the crossover. 
Then we solve the Josephson equations of the population imbalance and the relative phase 
calculating the frequency of tunneling oscillation both in the linear regime and in the nonlinear one. 
Our results show that the Josephson frequency is larger in the intermediate regime of the BCS-BEC crossover 
due to the peculiar behavior of the effective tunneling energy in the crossover. 
\end{abstract}

\pacs{03.75.Lm; 03.75.Ss; 67.85.-d}

\maketitle

\section{Introduction}

Recent developments in confinement, cooling, and control of interaction with alkali-metal atoms 
have renewed the interest on the BCS-BEC crossover. 
This crossover from the Bardeen-Cooper-Schrieffer (BCS) state to the Bose-Einstein condensate (BEC) has been observed in two-hyperfine-components 
Fermi vapours of $^{40}$K atoms and $^6$Li atoms 
\cite{Regal2004,Zwierlein2004,Kinast2004} with the use of Fano-Feshbach resonances \cite{Inouye1998}. 
Quite remarkably, the Josephson effect \cite{Josephson1962}, i.e. the macroscopic quantum tunneling 
of Cooper pairs \cite{Ambegaokar1963}, has been experimentally investigated in the 
BEC-BEC crossover \cite{Valtolina2015} with a three-dimensional (3D) configuration of neutral ultracold atoms. 
Moreover, also quasi two-dimensional (2D) ultracold Fermi gases 
have been realized \cite{Martiyanov2010} and studied in the 2D BCS-BEC 
crossover \cite{Makhalov2014, Ries2015, Boettcher2016, Fenech2016}.

In this paper we investigate theoretically the Josephson effect 
in the 2D BCS-BEC crossover by taking into account the crucial role of 
the condensate fraction and the chemical potential on the effective tunneling energy \cite{Zaccanti2019}. 
To obtain these quantities we solve the extended-BCS gap and number equations at zero temperature 
in the full 2D crossover. We analyze the effects of quantum fluctuations, comparing the results 
obtained applying mean-field and beyond mean-field approximations to the pairing field. 
Having a reliable beyond-mean-field effective tunneling energy, we solve the atomic 
Josephson junction equations for the population imbalance and the relative phase 
reported in Ref. \cite{Salasnich2007a} and adapted to the 2D case. In this way we obtain 
immediately the critical current of the direct-current Josephson effect,  
but also the oscillation frequency of the alternate-current Josephson effect, 
as a function of the interaction strength of the 2D BCS-BEC crossover.  
We also study the oscillation frequency of macroscopic quantum tunneling in the nonlinear 
regime finding that is higher with respect to the one of the linear regime. 

\section{2D BCS-BEC crossover}

We consider a 2D attractive Fermi gas of ultracold and dilute two-spin component neutral atoms. 
The Mermin-Wagner theorem \cite{Mermin1966, Coleman1973} says that for a system with  
spatial dimension $d=2$ there cannot be spontaneous symmetry breaking at finite temperature. 
In other words, there is a Bose-Einstein condensate only at zero temperature. 
Nonetheless, a 2D system can exhibit superfluidity both at zero and finite temperature:  
below the Berezinskii-Kosterlitz-Thouless (BKT) critical temperature $T_{BKT}$ \cite{Berezinsky1972, Kosterlitz1973}, 
there is superfluidity. The BKT phase transition is a topological phase transition with a jump of  
the superfluid density at $T_{BKT}$. As we shall see, in our 2D fermionic superfluid quantum 
and thermal fluctuations play a crucial role to describe accurately various properties of the system \cite{Bighin2016}. 

The Hamiltonian density of a fermionic system made of ultracold and dilute alkali-metal atoms is given by  
\begin{multline}
H=\overline{\Psi}_{\sigma}(\vec{r},t)\Bigl[-\frac{\nabla^2}{2m}-\mu\Bigr]\Psi_{\sigma}(\vec{r},t)\\
-g\overline{\Psi}_{\uparrow}(\vec{r},t)\overline{\Psi}_{\downarrow}(\vec{r},t)\Psi_{\uparrow}(\vec{r},t)\Psi_{\downarrow}(\vec{r},t)
\label{densham}
\end{multline}
where the $\Psi_{\sigma}(\vec{r},t)$ and $\overline{\Psi}_{\sigma}(\vec{r},t)$ are the complex Grassman fields with spins $\sigma=\uparrow,\downarrow$. This Hamiltonian density described a system with single-channel interaction where $g>0$ is the strength of the s-wave inter-atomic coupling. Contrary to the 3D case, in 2D attractive interatomic potentials exists a bound state for any value of the interaction strength {{$g$}} \cite{Bertaina2011, Marini1998}. In this way, it is possible to define the binding energy $\epsilon_B$, the energy that keeps the molecules together, in all the crossover. This latter can be written in terms of 2D fermionic scattering length $a_{2D}$ \cite{Mora2003} as 
\begin{equation}
\epsilon_B=\frac{4}{e^{2\gamma}}\frac{\hbar^2}{ma_{2D}^2}
\end{equation} 
where $\gamma=0.577..$ is the Euler-Mascheroni constant. The binding energy is conceptually more appealing than the scattering length, so in 2D case the crossover it is mapped using the ratio $\epsilon_B/\epsilon_F$, 
where $\epsilon_F=\hbar^2\pi n/m$ is the 2D Fermi energy of two-spin-coomponent 
non-interacting fermions with total number density $n$. 
It is possible to move from a BCS state of weakly-bound Cooper pairs to a BEC state of strongly-bound Cooper pairs by increasing $\epsilon_B$. The interaction strength $g$ is related to the binding energy $\epsilon_B$ by the expression
\begin{equation}
-\frac{1}{g}=\frac{1}{L^2}\sum_k\frac{1}{\epsilon_k+\frac{\epsilon_B}{2}}
\label{g2d}
\end{equation} 
where $\epsilon_k=\hbar^2k^2/2m$ is the single-particle energy. 

Through the Hubbard-Stratonovich transformation, the Hamiltonian density $\eqref{densham}$ can be rewritten 
introducing the bosonic complex field $\Delta(x)$ \cite{Altland2010}. 
The Hamiltonian density then becomes
\begin{equation}
H=\overline{\Psi}_{\sigma}\Bigl[-\frac{\nabla^2}{2m}-
\mu\Bigr]\Psi_{\sigma}+\frac{|\Delta|^2}{g}-\overline{\Delta}\Psi_{\downarrow}\Psi_{\uparrow}-
\Delta\overline{\Psi}_{\uparrow}\overline{\Psi}_{\downarrow}
\label{hs}
\end{equation}
Here we adopt a path integral approach \cite{Gorkov1959} and 
we introduce the action $S$ and the partition function $Z$ at temperature $T$ of the system:
\begin{gather}
S=\int_0^{\beta}d\tau\int d\vec{r}\Bigl[\overline{\Psi}_{\sigma}\partial_{\tau}\Psi_{\sigma}+H\Bigr]
\label{S}
\\
Z=\int_{}^{}\mathcal{D}[\Delta,\overline{\Delta}]\mathcal{D}[\Psi_{\sigma},\overline{\Psi}_{\sigma}]  exp(-{{S}}[\Delta, \overline{\Delta},\Psi_{\sigma},\overline{\Psi}_{\sigma}])
\label{Z}
\end{gather}
where $\beta=1/k_BT$ with $k_B$ Boltzmann constant. As we can see the density Hamiltonian $\eqref{hs}$ is quadratic 
with respect to $\Psi_\sigma$ so it is possible to integrate over it obtaining the effective action $S_{eff}[\Delta(\vec{r},t),\overline{\Delta}(\vec{r},t)]$. We want to study the effects of Gaussian fluctuation 
of the gap field $\Delta(\vec{r},t)$ and so we set:
\begin{equation}
\Delta(x)=\Delta_0+\eta(x)
\end{equation}
where $\eta(x)$ is the complex pairing field of bosonic fluctuations \cite{Nagaosa2013, Roberto2008}.  

First of all we investigate the mean-field approximation imposing $\Delta(x)=\Delta_0$, where $\Delta_0$ is a real 
and spatial independent value.  We replace it in $\eqref{S}$, $\eqref{Z}$ and obtaining the mean-field 
effective action $S_{mf}$ and partition function  
\begin{equation}
Z_{mf}=exp\Bigl[-\frac{S_{mf}}{\hbar}\Bigr]=exp\Bigl[{{\beta\Omega_{mf}}}\Bigr]
\end{equation}
where $\beta=1/(k_BT)$ and $\Omega_{mf}$ is the mean-field grand potential. 
At zero temperature ($T=0$, i.e. $\beta\rightarrow +\infty$) {{the 2D $\Omega_{mf}$ reads}} 
\begin{equation}
{{\Omega_{mf}=-\sum_k(E_k+\epsilon_k+\mu)-L^2\frac{\Delta_0^2}{g}}}
\end{equation}
{{where $E_k=\sqrt{(\epsilon_k-\mu)^2+\Delta_0^2}$}}.

To describe the BCS-BEC crossover we solve the 
mean-field gap and number equation of the system. 
The gap equation is calculated by applying the saddle-point condition 
\begin{equation}
\frac{\partial\Omega_{mf}}{\partial\Delta_0} = 0
\end{equation}
{{In this way, one obtains the familiar BCS gap equation}}
\begin{equation}
{{-\frac{1}{g}=\frac{1}{V}\sum_k\frac{1}{2E_k}}}
\label{gapbcs}
\end{equation}
{{which can be combined with Eq. (\ref{g2d}) 
to remove ultraviolet divergences and to get the 
energy gap $\Delta_0$ and the chemical potential $\mu$ 
as a fuction of the binding energy $\epsilon_B$.}} 

The mean-field number equation is instead derived by using the 
thermodynamic relation
\begin{equation}
n=-\frac{\partial\Omega_{mf}}{\partial\mu}
\label{n}
 \end{equation}
from which one obtains an implicit formula between the chemical 
potential $\mu$ and the number density $n$. 

{{By considering the fluctuation field 
$\eta(\vec{r},t)$ we find a new effective action, 
$S_{eff}=S_{mg}[\Delta_0]+S_g[\eta,\overline{\eta}]$, from which is 
possible to obtain the 2D grand potential $\Omega$ in the beyond mean-field 
approximation}}
\begin{equation}
{{\Omega=\Omega_{mf}+\Omega_g=\Omega_{mf}+
\frac{1}{2\beta}\sum_q\ln{det(\mathbb{M}(q))}}}
\label{gpbmf}
\end{equation}
{{where $\mathbb{M}(q)$ is the inverse pair 
fluctuation propagator, reported in the supplement material of 
Ref. \cite {Bighin2016}. It's important to note that 
the gap equation is independent on the approximation used. 
The relation $\eqref{gapbcs}$ is valid also in the beyond mean-field case. 
Instead the number equation depends on the approximation. 
Replacing the relation $\eqref{gpbmf}$ in the Eq.$\eqref{n}$ one obtains 
the beyond mean-field number equation
\begin{equation} 
n = - \frac{\partial\Omega}{\partial\mu} \;  
\label{n-bmf}
\end{equation}
}} 

The procedure of the the numerical calculations at the Gaussian (one-loop) 
level is detailed in Refs. \cite{Salasnich2014, Brendan2017}. 
Solving the gap equation (\ref{gapbcs}) and the beyond-mean-field 
number equation (\ref{n-bmf}) it is then possible to obtain the behavior of the 
single-particle chemical potential $\mu$ and energy gap 
$\Delta_0$ along the crossover. 

\begin{figure}[ht!]
\centering
\includegraphics[width=0.45\textwidth]{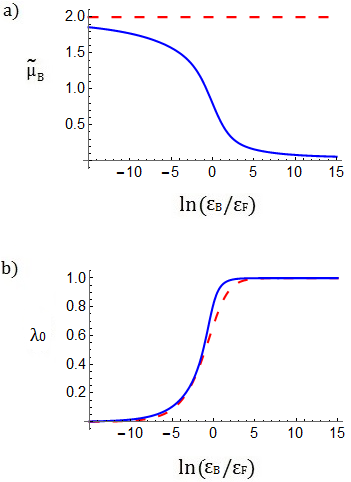}
 \caption{a) Mean-field (red dashed line) and beyond mean-field (black solid line) scaled bosonic 
chemical potential $\tilde{\mu}=\mu_B/\epsilon_F$ as a function of the scaled binding 
energy $\epsilon_B/\epsilon_F$. 
b) Mean-field (red dashed line) and beyond mean-field (black solid line) 
condensate fraction $\lambda_0= n_0/n$, where $n_0$ is the condensate pair density 
and $n$ is the total density, 
as a function of the scaled binding energy $\epsilon_B/\epsilon_F$. Here $\epsilon_F$ is the Fermi energy.}
\label{n0}
\end{figure}

In the panel a) of Fig. \ref{n0} we report the scaled chemical 
potential $\tilde{\mu}_B=\mu_B/\epsilon_F$, where $\mu_B=2\mu+\epsilon_B$ is the chemical potential of composite boson. In the mean-field case, $\mu_B$ is constant and equal to $2\epsilon_F$ in all the crossover. 
Gaussian fluctuations introduce a nontrivial crossover dependence. 
The effects are more relevant in the strong coupling limit (BEC regime) than in the weak coupling limit (BCS regime), like in the three-dimensional case \cite{Pascucci2020, Hu2006}. From BCS to BEC limit the Cooper pair size decreases until it becomes smaller than the average distance between the particles; at short distance fluctuations cannot be neglected. We have to go beyond the mean-field approximation to correctly describe the BCS-BEC crossover phenomenon. Another useful interesting quantity to study is the condensate density $n_0$, i.e. the density of the 
Cooper pairs with a vanishing center-of-mass linear momentum. 
We use Eq. (21) reported in the work \cite{Salasnich2007b} where the mean-field number of condensate couples $N_0$ is evaluated as the largest eigenvalue of the two-body density matrix, written through the Bogoliubov representation of the field operator $\psi_\sigma$
\begin{equation}
{{N_0=\int d^3r_1d^3r_2|\Psi_{\downarrow}(r_1)\Psi_{\uparrow}(r_2)|^2}}
\end{equation}
 The condensate fraction $\lambda_0$ of the 
fermionic system is then given by  
\begin{equation}
\lambda_0 = \frac{n_0}{n}=\frac{1}{2}\frac{\frac{\pi}{2}+arctan(\frac{\mu}{\Delta_0})}{\frac{\mu}{\Delta_0}+\sqrt{1+\frac{\mu^2}{\Delta_0^2}}}
\label{n0bmf}
\end{equation}
where $n_0$ is the condensate density of Cooper pairs and $n$ is the total fermionic density. 
Notice that here $n_0$ is defined such that $n_0\to n$ in the deep BEC regime. 
In Ref. \cite{Fukushima2007} Fukushima {\it et al.} 
calculated the effects of gaussian fluctuations on the condensate fraction for a 3D BCS-BEC crossover 
concluding that the Gaussian correction is quite small and it can be neglected 
in the full crossover. Here we evaluate the beyond-mean-field condensate fraction 
replacing in \eqref{n0bmf} the mean-field chemical potential and gap energy 
with their beyond mean-field counterparts \cite{Lianyi2015}. 
As we can see from the panel b) of Fig.\ref{n0}, in the deep BCS and BEC regime the two approximations give the same result, that is however quite obvious because the condensate fraction always starts from $0$ and goes to $1$. 
The main difference is found near the crossover point $\epsilon_B=\epsilon_F$. 

We look for the $\epsilon_B/\epsilon_F$ value over which the system is considered in the BEC regime. 
A Bose-Einstein condensate is characterized by a macroscopic number of particles in the ground-state energy. 
Following Ref. \cite{Guidini2014} we choose to consider the system in the strong-coupling BEC regime if the 80$\%$ of the Cooper pairs are condensed. In the beyond mean-field approximation, the condensate fraction reaches the value of 0.8 at the point $\ln{(\epsilon_B/\epsilon_F)}=-0.11$. We note that at $\ln(\epsilon_B/\epsilon_F)=-0.16$ the single-particle chemical potential $\mu$ becomes negative. Considering the assumptions made to obtain the beyond mean-field $n_0/n$, we can also say that the system enters the BEC regime when the single-particle chemical potential $\mu$ becomes negative. 

\section{Josephson effect}

The Josephson effect was introduced by Brian Josephson in 1962 \cite{Josephson1962}. 
He predicted that between two superconductors separated by an insulating layer a supercurrent, 
made of Cooper pairs, can flows. The same hypothesis can be applied also to neutral superfluid 
fermionic atoms replacing the insulating layer by a potential barrier created with a laser beam. 
Here we investigate the problem of a double-well potential and we consider the Fermi gas of $N$ atoms with two equally populated spin components and attractive inter-atomic strength at zero temperature. The system can be schematized considering two reservoirs, a left (L) one and a right (R) one, which contain superfluid particles with $A/2$ the area of each 2D reservoirs. 
Following the procedure of Ref.\cite{Salasnich2007a}, it is possible to describe the time evolution of the system by introducing the two-state phenomenological model 
\begin{gather}
i\hbar\frac{\partial} {\partial t}\Psi_L(t)=E_L \, \Psi_L(t)-K\, \Psi_R(t)
\label{eqL}
\\
i\hbar\frac{\partial}{\partial t}\Psi_R(t)=E_R \, \Psi_R(t)-K\, \Psi_L(t)
\label{eqR}
\end{gather} 
where $\Psi_\alpha(t)$ is the time-depdendent macroscopic wavefunction 
in the 2D reservoir $\alpha=L,R$ and $E_{\alpha} = 2 \mu(2|\Psi_{\alpha}(t)|^2/A)$ 
is the onsite energy of that reservoir. 
The hopping term $K$ describes the effective tunneling energy between the two regions. 
It is quite common to use a phenomenological tunneling energy to describe Josephson effect in superconductors because a microscopic analytic derivation of $K$, valid in all the crossover, 
is not yet available. Under the assumption that the potential barrier keeps 
the particles between the two reservoirs weakly-linked, we can write
\begin{equation}
\Psi_{\alpha}(t)=\sqrt{N_{\alpha}(t)/2}\exp(i\theta_{\alpha}(t))
\label{psit}
\end{equation}
where $N_{\alpha}(t)$ and $\theta_{\alpha}(t)$ are the number of fermions and the 
superfluid phase in the reservoir $\alpha$ at the time $t$. 
To describe the Josephson effect we introduce the relative phase 
\begin{equation}
\phi(t)=\theta_R(t)-\theta_L(t)
\label{pd}
\end{equation}
and the adimensional population imbalance 
\begin{equation}
z(t)=\frac{N_L(t)-N_R(t)}{N}
\label{rnb}
\end{equation}
Replacing $\eqref{psit}$, $\eqref{pd}$ and $\eqref{rnb}$ in $\eqref{eqL}$, $\eqref{eqR}$ one obtains 
\begin{eqnarray}
\dot{z}&=&-\frac{2K}{\hbar}\sqrt{1-z^2}\sin(\phi)
\label{jep}
\\
\dot{\phi}&=&\frac{2}{\hbar}\Bigr[{{\mu_1+\mu_2}}\Bigr] 
+\frac{2K}{\hbar}\frac{z}{\sqrt{1-z^2}}\cos(\phi)
\label{jez}
\end{eqnarray}
where {{$\mu_1$ and $\mu_2$ are the chemical potential corrisponding 
to the densities n(1+z) and n(1-z) respectively}}. 
Eqs. \eqref{jep} and \eqref{jez} are the atomic Josephson junction equations  
for the two dynamical variable $z(t)$ and $\phi(t)$. 
They describe the oscillations of $N$ fermionic atoms 
that tunnel between the left region L and the right region R. 
A crucial role is played by the chemical potential $\mu(n)$, that is obtained 
by solving self-consistently the gap equation and the number equation. 
The macroscopic tunneling can be described as a tunneling current defined in the following way
\begin{equation}
 I=-\dot{z}N/2=\Bigl(\frac{KN}{\hbar}\Bigr)\sqrt{1-z^2} \sin(\phi)=I_c\sqrt{1-z^2} \sin(\phi)
\end{equation}
where $I_c=KN/\hbar$ is the critical current of the direct-current Josephson effect 
where $\phi$ is time independent. 
Thus, we need  a coupling-dependent formula for the effective tunneling energy $K$ 
to get the critical current $I_c$, and also to study the alternate-current Josephson effect 
which is obtained by solving Eqs. (\ref{jep}) and (\ref{jez}) without restrictions 
on the dynamics of the relative phase. 

\section{Coupling-dependent tunneling energy}

In this Section we consider a time-independent phase difference ($\phi(t)=\overline{\phi}$) and a small population 
imbalance $|z(t)|\ll 1$.  In this case the Josephson current $I$ reads 
\begin{equation}
I\Bigl(\overline{\phi}\Bigr)=I_c \sin\Bigl(\overline{\phi}\Bigr) 
\end{equation}
This is the familiar formula of the direct-current Josephson effect \cite{Josephson1962, Ambegaokar1963}. 

In a recent work, Zaccanti and Zwerger \cite{Zaccanti2019} developed a model to describe 3D Josephson tunneling between two reservoirs of ultracold superfluid atoms which account the dependence of the critical current on the coupling. 
In particular, they found the critical current density {{$j_c$, such that $I_c=Aj_c$ where A is the transverse area of the Josephson junction}}, is given by 
\begin{equation}
\frac{j_c}{|t|(\mu_B)}=\frac{j_c}{|t|^2(\mu_F)}=\frac{nv_F}{8}\lambda_0\sqrt{\tilde{\mu}}
\label{zacc}
\end{equation}
where $\lambda_0=n_0/n$ is the condensate fraction, $\tilde{\mu}=\mu_B/2\epsilon_F$ is the renormalized bosonic chemical potential and $v_F=\sqrt{2\epsilon_F/m}$ is the Fermi velocity of the non-interacting Fermi gas. 
The $|t|(\mu_B)$ and $|t|^2(\mu_F)$, where $\mu_F=\mu_B/2$, are the associated single-boson transmission 
amplitude and the transmission probability of a single fermion respectively. 
Eq. (\ref{zacc}) provides a very useful formula valid for the BCS-BEC crossover both in 3D. {{This result has been obtained evaluating the tunneling amplitude like the problem is 1D, since the potential barrier extends only in one direction \cite{Meier2001}. We are considering the same geometry for the potential barrier and for this reason we think that the results obtained by Zaccanti and Zwerger can be extended also in the 2D systems.}} 

\begin{figure}[ht!]
\centering
\includegraphics[width=0.45\textwidth]{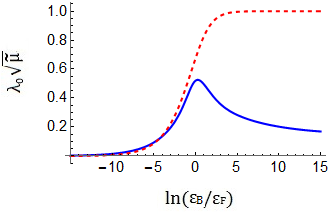}
\caption{The mean-field (red dashed curve) and beyond mean-field 
(blue solid curve) $\lambda_0\sqrt{\tilde{\mu}}$ as a function 
of the scaled binding energy $\epsilon_B/\epsilon_F$, where $\lambda_0=n_0/n$ 
is the ratio between the condensate density and 
the total density, $\tilde{\mu}=\mu_B/2\epsilon_F$ is the ratio between 
the bosonic chemical potential and two times the Fermi  energy $\epsilon_F$. }
 \label{k2d}
\end{figure}

Under the condition of a high potential barrier both $|t|(\mu_B)$ and $|t|^2(\mu_F)$ are practically 
constant along the crossover and the dependence on the coupling strength is mainly in the product $\lambda_0\sqrt{\tilde{\mu}}$. 
Its behavior along the 2D BCS-BEC crossover is shown in Fig. $\ref{k2d}$ both with mean-field and beyond-mean-field schemes.  
The figure shows that the mean-field curve grows monotonically. Instead, 
the beyond mean-field curve grows from the BCS regime to a maximum at the value 
$\ln(\epsilon_B/\epsilon_F)=0.26$ and 
then it decreases in the BEC regime. Thus, also in this case the quantum fluctuations cannot be neglected 
in the strong-coupling regime. To better understand the meaning of this product we can write Eq. \eqref{zacc} 
in the form of the critical current $\hbar I_c=KN$ where K is the tunneling energy and N the total number of fermions.
\begin{equation}
\hbar I_c=\hbar L_1 j_c=\hbar L_1 |t|^2(\mu_F) \frac{nv_F}{8}\lambda_0 \sqrt{\tilde{\mu}}
\end{equation}   
Replacing $n=N/(2{{L_1L_2}})$, where {{$L_1$ and $L_2$}} are the 
trasverse and longitudinal lengths of the 2D system, one obtains
\begin{equation}
\hbar I_c=\hbar|t|^2(\mu_F) \frac{Nv_F}{16L_2}\lambda_0 \sqrt{\tilde{\mu}}
\end{equation}
Comparing with $\hbar I_c=KN$, the tunneling energy $K$ reads
\begin{equation}
K= \hbar|t|^2(\mu_F) \frac{v_F}{16L_2}\lambda_0 \sqrt{\tilde{\mu}}
= K_0 \lambda_0 \sqrt{\tilde{\mu}}
\label{Kcd}
\end{equation}
where the factor $K_0=\hbar|t|^2(\mu_F)v_F/16L_2$ encloses all the coupling 
independent factors. Once the system setup is fixed,$K_0$ is a constant 
along the crossover. So the product $\lambda_0 \sqrt{\tilde{\mu}}$ provides 
useful informations about the behavior of the tunneling energy and the 
critical current along the crossover.

\section{Josephson dynamics of 2D fermionc superfluids}

We replace the coupling dependent tunneling energy \eqref{Kcd} in the Josephson equations \eqref{jep} and 
\eqref{jez} to study the time evolution of the population imbalance $z(t)$ 
and the relative phase $\phi(t)$. The Josephson junction equations become
\begin{eqnarray}
\dot{z}&=&-\frac{2K_0\lambda_0\sqrt{\tilde{\mu}}}{\hbar}\sqrt{1-z^2}\sin(\phi)
\label{jepk}
\\
\dot{\phi}&=&\frac{2}{\hbar}\Bigr[\mu\Bigl(n(1+z)\Bigr)+\mu\Bigl(n(1-z)\Bigr)\Bigr] 
\nonumber 
\\
&+&\frac{2K_0\lambda_0\sqrt{\tilde{\mu}}}{\hbar}\frac{z}{\sqrt{1-z^2}}\cos(\phi)
\label{jezk}
\end{eqnarray}
Since in the previous sections we show in detail the crucial effect of the quantum fluctuations, 
in this part we treat the system with the beyond mean-field approximation directly. 

\begin{figure}[ht!]
\centering
\includegraphics[width=0.4\textwidth]{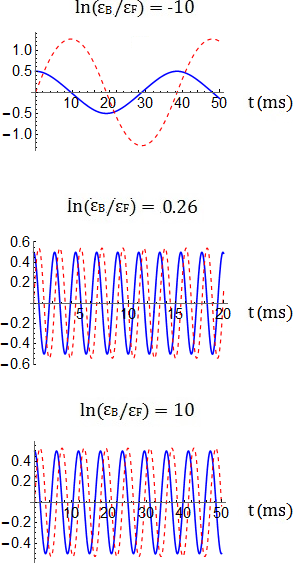}
\caption{Time evolution of the phase difference $\phi(t)$ in $[rad]$ 
(red dashed curve) and population 
imbalance $z(t)$ (blue solid curve) for three values of 
the scaled binding energy $\epsilon_B/\epsilon_F$. 
The numerical results are obtained solving Eqs. (\ref{jepk}) and (\ref{jezk}) 
with the initial conditions $z(0)=0.5$ and $\phi(0)=0$.}
 \label{eb}
\end{figure}

We choose to study a fermionic superfluid of $^{40}K$ atoms with the total density $n=0.01$ atoms/$\mu m^2$ 
and $K_0/k_B=2\cdot10^{-8}$ Kelvin. In Fig. \ref{eb} we show the 
time evolution of $z(t)$ and $\phi(t)$ for three different crossover 
points: $\ln(\epsilon_B/\epsilon_F)= -10$, 
$\ln(\epsilon_B/\epsilon_F)= 10$ and $\ln(\epsilon_B/\epsilon_F)= 0.26$. 

The oscillation frequency of both quantities changes along the crossover. 
We note anyway that $z(t)$ and $\phi(t)$ oscillate with the same 
frequency $\omega$ at a fixed scaled binding energy $\epsilon_B/\epsilon_F$. 

\begin{figure}[htb] 
\centering
\includegraphics[width=0.4\textwidth]{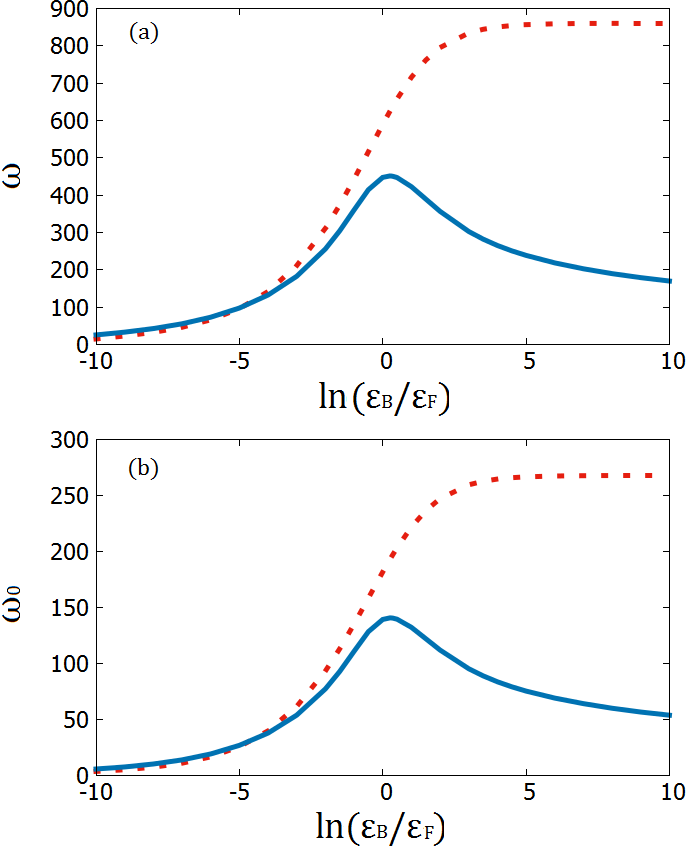}
\caption{(a) Josephson oscillation frequency $\omega$ in the nonlinear regime 
as a function of the scaled binding energy $\epsilon_B/\epsilon_F$. 
The frequency $\omega$ is obtained by solving Eqs. \eqref{jepk} 
and \eqref{jezk} with initial conditions: $z(0)=0.5$ and $\phi(0)=0$. 
(b) Josephson oscillation frequency $\omega_0$ in the linear regime 
as a function of the scaled binding energy $\epsilon_B/\epsilon_F$. 
The frequency $\omega_0$ is obtained from Eq. \eqref{omega0} 
In the tunneling energy $K$ of Eq. (\ref{Kcd}) we use 
$K_0=2\cdot10^{-8}k_B$ Kelvin. 
Red dashed curves: mean-field approximation; blue solid curves: 
beyond-mean-field (Gaussian) results.
Here $\epsilon_F$ is the Fermi energy.}
\label{osc}
\end{figure}

We determine the frequency $\omega$ by evaluating the time interval 
between an oscillation peak and the next one. 
In the upper panel of Fig. \ref{osc} we report the Josephson oscillation 
frequency $\omega$ obtained by solving Eqs. (\ref{jepk}) and (\ref{jezk}) 
with the initial conditions $z(0)=0.5$ and $\phi(0)=0$. In the panel we 
plot both mean-field (red dashed curve) and beyond-mean-field 
(blue solid curve) results.

We also linearize the Josephson equations. The linearized Josephson 
equations admit a stable stationary solution with 
$\overline{z}=0$ and $\overline{\phi}=2\pi j$, for integer j. 
The linear oscillations around this stable solution with $j=0$ are 
charactedized by the frequency 
\begin{equation}
\omega_0=\frac{K_0 \lambda_0 \sqrt{\tilde\mu}}{\pi\hbar}
\sqrt{1+\frac{2mc_s^2}{K_0\lambda_0\sqrt{\tilde\mu}}}
\label{omega0}
\end{equation}
where $c_s=\sqrt{(n/m)\partial\mu/\partial n}$ is the speed of sound. 
This frequency $\omega_0$ is called zero-mode. Notice that also here 
we have replaced the coupling dependent relation $\eqref{Kcd}$ in the 
tunneling energy $K$. Fig. \ref{osc} shows that $\omega$ is always 
larger than $\omega_0$ and their behavior as a function 
of $\ln(\epsilon_B/\epsilon_F)$ is very similar to the product 
$\lambda_0\sqrt{\tilde{\mu}}$ both in mean-field approximation 
(red dashed curves) and in beyond mean-field one (blue solid 
curves).  

\section{Conclusions}

The phenomenological Josephson equations for the relative phase and the population imbalance 
depend on two effective coupling parameters: the tunneling energy and the onsite energy. 
As recently shown by Zaccanti and Zwerger \cite{Zaccanti2019}, in general these parameters 
are a function of the chemical potential and the condensate fraction of the system. 
For a two-dimensional fermionic superfluid we have derived all these quantities at zero temperature 
from a microscopic approach based on functional integration, finding that the Gaussian 
quantum fluctuations of the pairing field strongly modify the results in the BEC regime 
of the BCS-BEC crossover. By using the beyond-mean-field chemical potential 
and tunneling energy we have then numerically and analytically studied the Josephson equations. 
We have found that the critical current of the direct-current Josephson effect and the 
frequency of the alternate-current Josephson effect are larger in the intermediate 
regime of the BCS-BEC crossover because they are both proportional 
the effective tunneling energy which has the same behavior. 
The Josephson effect has been observed with fermionic superfluids made of alkali-metal atoms 
in a three-dimensional configuration of the BCS-BEC crossover \cite{Valtolina2015} 
and very recently also with two-dimensional bosonic atoms \cite{unknown}. 
We believe that our theoretical results can be a useful benchmark for the next future experiments 
on the macroscopic quantum tunneling with ultracold atoms in the two-dimensional BCS-BEC crossover. 

\section*{Acknowledgments}

The authors acknowledge Matteo Zaccanti and Wilhelm Zwerger for useful e-suggestions. 
LS thanks the BIRD project "Time-dependent density functional theory of quantum atomic mixtures"  
of the University of Padova for partial support.

\end{document}